\documentclass[11pt]{article}
\usepackage{graphicx}
\usepackage{longtable,lscape}
\usepackage{amsthm}
\usepackage{amsfonts}
\usepackage{amsmath}
\usepackage{bbm}
\usepackage{float}
\usepackage{url}
\newtheorem{definition}{Definition}
\newtheorem{theorem}{Theorem}

\newcommand{\captionfonts}{\footnotesize}
\makeatletter  
\long\def\@makecaption#1#2{%
  \vskip\abovecaptionskip
  \sbox\@tempboxa{{\captionfonts #1: #2}}%
  \ifdim \wd\@tempboxa >\hsize
    {\captionfonts #1: #2\par}
  \else
    \hbox to\hsize{\hfil\box\@tempboxa\hfil}%
  \fi
  \vskip\belowcaptionskip}
\makeatother 
\begin{document}
\title{General Quantum Hilbert Space Modeling Scheme for Entanglement}
\author{Diederik Aerts and Sandro Sozzo \vspace{0.5 cm} \\ 
        \normalsize\itshape
        Center Leo Apostel for Interdisciplinary Studies \\
        \normalsize\itshape
        and, Department of Mathematics, Brussels Free University \\ 
        \normalsize\itshape
         Krijgskundestraat 33, 1160 Brussels, Belgium \\
        \normalsize
        E-Mails: \url{diraerts@vub.ac.be,ssozzo@vub.ac.be}
           \\
              }
\date{}
\maketitle
\begin{abstract}
\noindent
We work out a classification scheme for quantum modeling in Hilbert space of any kind of composite entity violating Bell's inequalities and exhibiting entanglement. Our theoretical framework includes situations with entangled states and product measurements (`customary quantum situation'), and also situations with both entangled states and entangled measurements (`nonlocal box 
 situation', `nonlocal non-marginal box situation'). We show that entanglement is structurally a joint property of states and measurements. Furthermore, entangled measurements enable quantum modeling of situations that are usually believed to be `beyond quantum'. Our results are also extended from pure states to quantum mixtures.
\end{abstract}

\section{Introduction\label{intro}}
Entanglement is one of the most intriguing aspects of quantum physics. Entailing the violation of 
 `Bell's inequalities', it is responsible for a number of non-classical and far from well understood phenomena, such as quantum nonlocality \cite{bell1964} and quantum non-Kolmogorovness \cite{af1982,aerts1986,pitowsky1989}. Several powerful potential applications of entanglement have been identified, which make it one of the most important study objects in quantum information theory and quantum technology. Otherwise impossible tasks, such as `superdense coding' and `teleportation', the `quantum key distribution' and other protocols in quantum cryptography, the basic algorithms in quantum computation, exploit entanglement and its basic features. And, more, advanced experimental techniques, such as ion trapping and some fundamental processes in quantum interferometry, need entanglement and its purification, characterization and detection.

What was additionally very amazing is that entanglement appears, together with some other quantum features (contextuality, emergence, indeterminism, interference, superposition, etc.), also outside the microscopic domain of quantum theory, in the dynamics of concepts and decision processes within human thought, in computer science, in biological interactions, etc. These
results constituted the beginning of a systematic and promising search for quantum structures and the 
use of quantum-based models in domains where classical structures show to be problematical \cite{aertsaerts1995,aertsgabora2005a,aertsgabora2005b,aerts2009,pb2009,k2010,bb2012}. Coming to our research, many years ago we already identified situations in macroscopic physics which violate Bell's inequalities \cite{aerts1982,aertsaertsbroekaertgabora2000}. More recently, we have performed a cognitive test showing that a specific combination of concepts violates Bell's inequalities \cite{as2011,ags2012,abgs2012}. These two situations explicitly exhibit entanglement and present deep structural and conceptual analogies \cite{asIQSA2012,1asQI2013,2asQI2013}. 
Resting on these findings, in the present paper, we put forward a general analysis and elaborate a global framework for the mathematical description of (not necessarily physical) composite entities violating Bell's inequalities. The entanglement in these situations is detected, represented and classified within explicit quantum models, where states, measurements and probabilities are expressed in the typical Hilbert space formalism. Our quantum-theoretic approach identifies different types of situations according to the quantum description that is required for their modeling.    

(i) Bell's inequalities are violated within `Tsirelson's bound' \cite{tsirelson80} and the marginal distribution law holds (`customary quantum situation'). In this case, entangled states and product measurements are present.

(ii) Bell's inequalities are violated within Tsirelson's bound and the marginal distribution law is violated (`nonlocal non-marginal box situation 1'). In this case, both entangled states and entangled measurements are present.

(iii) Bell's inequalities are violated beyond Tsirelson's bound and the marginal distribution law is violated (`nonlocal non-marginal box situation 2'). In this case, both entangled states and entangled measurements are present.

(iv) Bell's inequalities are violated beyond Tsirelson's bound and the marginal distribution law holds (`nonlocal box situation'). In this case, both entangled states and entangled measurements  are present.

Technical aspects, definitions and results will be introduced in Secs. \ref{qmath} (pure states) and \ref{mixtures} (mixtures). We will show that entanglement is generally a joint feature of states and measurements. If only one measurement is at play and the situation of a pure state is considered, entanglement is identified by factorization of probabilities and can be distributed between state and measurement. If more measurements are at play, the violation of Bell's inequalities is sufficient to reveal entanglement in both the 
pure 
and the 
mixed
case. But, in both cases the marginal distribution law imposes serious constraints in the ways this entanglement can be distributed.

Cases (i) are the customary situations considered in Bell-type experiments on microscopic quantum particles, but they are very special in our analysis and approach. They correspond to situations where the symmetry of the entity is such that all the entanglement of the situation can be pushed into the state, allowing a model with only product measurements, where the marginal distribution law is satisfied. Cases (ii) instead seem to be present in situations of real quantum spin experiments (a mention of the `experimental anomaly' indicating in our opinion the presence of entangled measurements occurs already in Alain Aspect PhD thesis \cite{aspect1982}, and was identified more explicitly in \cite{ak}). Case (iii) will be studied in Sec. \ref{nonlocalnonmarginalbox}, where we observe that a quantum model can be worked out also for situations beyond Tsirelson's bound, contrary to widespread beliefs. Finally, in Sec. \ref{nonlocalbox} we put forward an example of case (iv), namely, the so-called `nonlocal box', which is studied as a purely theoretical construct -- no physical realizations have been found prior than the ones we present here  -- in the foundations of quantum theory \cite{rp94}.

\section{Experiments on composite entities\label{experiments}}
In this section, we introduce the experimental setup we aim to represent in our formalism. Let $S$ be a composite entity made up of the sub-entities $S_A$ and $S_B$, and let $S$ be prepared in the state $p$. A `Bell-type experimental setting' can be described as follows.
 
We denote the single (dichotomic) measurements on $S_A$ and $S_B$ by $e_{A}$, $e_{A'}$, respectively, $e_{B}$, $e_{B'}$, with outcomes 
$\lambda_{A_{1}}$, $\lambda_{A_{2}}$, $\lambda_{A'_{1}}$, $\lambda_{A'_{2}}$, $\lambda_{B_{1}}$, $\lambda_{B_{2}}$ and $\lambda_{B'_{1}}$, $\lambda_{B'_{2}}$, respectively. Let us consider $e_A$. If the outcome $\lambda_{A_{1}}$ ($\lambda_{A_{2}}$) is obtained for $e_A$, then the state $p$ changes into a state $p_{A_{1}}$ ($p_{A_{2}}$). Analogously, we can associate final states with the other measurements.

We denote the coincidence measurements on $S$ by $e_{AB}$, $e_{AB'}$, $e_{A'B}$ and $e_{A'B'}$, which involve both $S_A$ and $S_B$ ($e_{AB}$ can, e.g., be performed by performing $e_A$ on $S_A$ and $e_B$ on $S_B$, but it can be a more general measurement). The measurement $e_{AB}$ has four outcomes $\lambda_{A_{1}B_{1}}$, $\lambda_{A_{1}B_{2}}$, $\lambda_{A_{2}B_{1}}$ and $\lambda_{A_{2}B_{2}}$, and four final states $p_{A_{1}B_{1}}$, $p_{A_{1}B_{2}}$, $p_{A_{2}B_{1}}$ and $p_{A_{2}B_{2}}$. The measurement $e_{AB'}$ has four outcomes $\lambda_{A_{1}B'_{1}}$, $\lambda_{A_{1}B'_{2}}$, $\lambda_{A_{2}B'_{1}}$ and $\lambda_{A_{2}B'_{2}}$, and four final states $p_{A_{1}B'_{1}}$, $p_{A_{1}B'_{2}}$, $p_{A_{2}B'_{1}}$ and $p_{A_{2}B'_{2}}$. The measurement $e_{A'B}$ has four outcomes $\lambda_{A'_{1}B_{1}}$, $\lambda_{A'_{1}B_{2}}$, $\lambda_{A'_{2}B_{1}}$ and $\lambda_{A'_{2}B_{2}}$, and four final states $p_{A'_{1}B_{1}}$, $p_{A'_{1}B_{2}}$, $p_{A'_{2}B_{1}}$ and $p_{A'_{2}B_{2}}$. Finally, the measurement $e_{A'B'}$ has four outcomes $\lambda_{A'_{1}B'_{1}}$, $\lambda_{A'_{1}B'_{2}}$, $\lambda_{A'_{2}B'_{1}}$ and $\lambda_{A'_{2}B'_{2}}$, and four final states $p_{A'_{1}B'_{1}}$, $p_{A'_{1}B'_{2}}$, $p_{A'_{2}B'_{1}}$ and $p_{A'_{2}B'_{2}}$.

Let us now consider the coincidence measurement $e_{AB}$. In it, the outcomes $\lambda_{A_{1}B_{1}}$, $\lambda_{A_{1}B_{2}}$, $\lambda_{A_{2}B_{1}}$ and $\lambda_{A_{2}B_{2}}$ are respectively associated with the probabilities $p(\lambda_{A_{1}B_{1}})$, $p(\lambda_{A_{1}B_{2}})$, $p(\lambda_{A_{2}B_{1}})$ and $p(\lambda_{A_{2}B_{2}})$ in the state $p$. 
We define in a standard way the expectation value $E(A,B)$ for the measurement $e_{AB}$ as $E(A,B)=p(\lambda_{A_{1}B_{1}})+p(\lambda_{A_{2}B_{2}})-p(\lambda_{A_{1}B_{2}})-p(\lambda_{A_{2}B_{1}})$,
hence considering $\lambda_{A_{1}B_{1}}=\lambda_{A_{2}B_{2}}=+1$ and $\lambda_{A_{1}B_{2}}=\lambda_{A_{2}B_{1}}=-1$. Similarly, we define the expectation values $E(A,B')$, $E(A',B)$ and $E(A',B')$ in the state $p$ for the coincidence measurements $e_{AB'}$, $e_{A'B}$ and $e_{A'B'}$, respectively. 

Finally, we introduce the quantity $\Delta=E(A',B')+E(A,B')+E(A',B)-E(A,B)$, and the Clauser-Horne-Shimony-Holt (CHSH) version of Bell's inequalities, that is, $-2\le\Delta\le 2$ \cite{chsh69}. If ${\mathcal S}_{p}$ is a set of experimental data on the entity $S$ in the state $p$ for the measurements $e_{AB}$, $e_{AB'}$, $e_{A'B}$ and $e_{A'B'}$ and the CHSH inequality is satisfied, then a single probability space exists for ${\mathcal S}_{p}$ which satisfies the axioms of Kolmogorov (classical, or `Kolmogorovian', probability). If the CHSH inequality is violated, we say that entanglement occurs between $S_A$ and $S_B$, since such a classical space does not exist in this case \cite{af1982,aerts1986,pitowsky1989}.

\subsection{Entanglement and pure states\label{qmath}}
Let us now come to a quantum-mechanical representation of the situation in Sec. \ref{experiments} in the Hilbert space ${\mathbb C}^{2} \otimes {\mathbb C}^{2}$, canonically isomorphic to ${\mathbb C}^{4}$ by means of the correspondence $|1,0,0,0\rangle \leftrightarrow |1,0\rangle\otimes|1,0\rangle$, $|0,1,0,0\rangle \leftrightarrow |1,0\rangle\otimes |0,1\rangle$, $|0,0,1,0\rangle \leftrightarrow |0,1\rangle\otimes|1,0\rangle$, $|0,0,0,1\rangle \leftrightarrow|0,1\rangle\otimes|0,1\rangle$, where $\{ |1,0\rangle, |0,1\rangle\}$ and $\{|1,0,0,0\rangle,|0,1,0,0\rangle,|0,0,1,0\rangle,|0,0,0,1\rangle\}$ are the canonical bases of ${\mathbb C}^{2}$ and ${\mathbb C}^{4}$, respectively. Let us also recall that the vector space ${\mathcal L}({\mathbb C}^{4})$ of all linear operators on ${\mathbb C}^4$ is isomorphic to the vector space  ${\mathcal L}({\mathbb C}^{2}) \otimes {\mathcal L}({\mathbb C}^{2})$, where ${\mathcal L}({\mathbb C}^{2})$ is the vector space of all linear operators on ${\mathbb C}^{2}$. The canonical isomorphism above introduces hence a corresponding canonical isomorphism between ${\mathcal L}({\mathbb C}^{4})$ and ${\mathcal L}({\mathbb C}^{2}) \otimes {\mathcal L}({\mathbb C}^{2})$. Both these isomorphisms will be denoted by $\leftrightarrow$ in the following, and we will work in both spaces ${\mathbb C}^{4}$ and ${\mathbb C}^{2} \otimes {\mathbb C}^{2}$ interchangeable.

Let us put forward a completely general quantum representation, where the state $p$ is represented 
by the unit vector $|p\rangle \in {\mathbb C}^{4}$,
and the measurement $e_{AB}$ by the spectral family constructed on the ON basis $ \{ |p_{A_{1}B_{1}}\rangle, |p_{A_{1}B_{2}}\rangle , |p_{A_{2}B_{1}}\rangle, |p_{A_{2}B_{2}}\rangle \}$, 
where the unit vector $|p_{A_{i}B_{j}}\rangle$ represents the state $p_{A_{i}B_{j}}$, $i,j=1,2$. Hence, $e_{AB}$ is represented by the self-adjoint operator ${\mathcal E}_{AB}=\sum_{i,j=1}^{2}\lambda_{A_{i}B_{j}}|p_{A_{i}B_{j}}\rangle \langle p_{A_{i}B_{j}}|$. Analogously, we can construct the self-adjoint operators ${\mathcal E}_{AB'}, {\mathcal E}_{A'B}, {\mathcal E}_{A'B'}$ respectively representing $e_{AB'}$, $e_{A'B}$ and $e_{A'B'}$. 

The probabilities of the outcomes of $e_{AB}$, $e_{AB'}$, $e_{A'B}$ and $e_{A'B'}$ in the state $p$ are respectively given by $p(\lambda_{A_{i}B_{j}})=|\langle p_{A_{i}B_{j}}|p\rangle|^{2}$, $p(\lambda_{A_{i}B'_{j}})=|\langle p_{A_{i}B'_{j}}|p\rangle|^{2}$, $p(\lambda_{A'_{i}B_{j}})=|\langle p_{A'_{i}B_{j}}|p\rangle|^{2}$ and $p(\lambda_{A'_{i}B'_{j}})=|\langle p_{A'_{i}B'_{j}}|p\rangle|^{2}$.

Moreover, if we put $\lambda_{X_{i}Y_{i}}=+1$, $\lambda_{X_{i}Y_{j}}=-1$, $i,j=1,2$, $j \ne i$, $X=A,A'$, $Y=B,B'$, we can write the expectation value in the state $p$ as $E(X,Y)=\langle p|{\mathcal E_{XY}}|p\rangle$ and the Bell operator by $B={\mathcal E}_{AB'}+{\mathcal E}_{A'B}+{\mathcal E}_{AB'}-{\mathcal E}_{AB}$. Thus, the CHSH inequality can be written as $-2 \le \langle p|B|p\rangle \le +2$.

Let us now introduce the notions of `product state' and `product measurement'. 
\begin{definition}
A state $p$, represented by the unit vector $|p\rangle \in {\mathbb C}^4$, is a `product state' if there exists two states 
$p_A$ and 
$p_B$, represented by the unit vectors 
$|p_A\rangle \in {\mathbb C}^2$ and 
$|p_B\rangle \in {\mathbb C}^2$, respectively, such that $|p\rangle \leftrightarrow 
|p_A\rangle\otimes|p_B\rangle$. Otherwise, $p$ is an `entangled state'.
\end{definition}
\begin{definition}
A measurement $e$, represented by a self-adjoint operator ${\cal E}$ in ${\mathbb C}^4$, is a `product measurement' if there exists
measurements 
$e_A$ and 
$e_B$, represented by the self-adjoint operators  
${\cal E}_A$ and 
${\cal E}_B$, respectively, in ${\mathbb C}^2$, such that ${\cal E} \leftrightarrow 
{\cal E}_A \otimes {\cal E}_B$. Otherwise, $e$ is an `entangled measurement'.
\end{definition}
Let now $p$ be a product state, represented by 
$|p_A\rangle\otimes|p_B\rangle$, where 
$|p_A\rangle$ and 
$|p_B\rangle$ represent the states 
$p_A$ and 
$p_B$, respectively. And let $e$ be a product measurement, represented by 
${\mathcal E}_A \otimes {\mathcal E}_B$, where 
${\cal E}_A$ and 
${\cal E}_B$ represent the measurements 
$e_A$ and 
$e_B$, respectively. The following theorems hold \cite{asIQSA2012}.
\begin{theorem}\label{th1}
The spectral family of the self-adjoint operator 
${\cal E}_A \otimes {\cal E}_B$ representing the product measurement $e$ has the form 
$|p_{A1}\rangle \langle p_{A1}|\otimes |p_{B1}\rangle \langle p_{B1}|$, $|p_{A1}\rangle \langle p_{A1}| \otimes |p_{B2}\rangle \langle p_{B2}| $, $|p_{A2}\rangle \langle p_{A2}| \otimes |p_{B1}\rangle \langle p_{B1}|$ and $|p_{A2}\rangle \langle p_{A2}| \otimes |p_{B2}\rangle \langle p_{B2}|$,
where 
$|p_{A1}\rangle \langle p_{A1}|$ and 
$|p_{A2}\rangle \langle p_{A2}|$ is a spectral family of 
${\cal E}_A$ and 
$|p_{B1}\rangle \langle p_{B1}|$ and 
$|p_{B2}\rangle \langle p_{B2}|$ is a spectral family of 
${\cal E}_B$.
\end{theorem}
Theorem \ref{th1} states that the spectral family of a product measurement is constructed on an ON basis of product states.
\begin{theorem} \label{th2}
Let $p$ be a product state represented by 
$|p_A\rangle \otimes |p_B\rangle$, and $e$ a product measurement represented by 
${\cal E}_A \otimes {\cal E}_B$. Then, there exists probabilities $p(\lambda_{A_{1}})$, $p(\lambda_{B_{1}})$, $p(\lambda_{A_{2}})$ and $p(\lambda_{B_{2}})$, where $p(\lambda_{A_{i}})$ ($p(\lambda_{B_{i}})$) is the probability for the outcome $\lambda_{A_{i}}$ ($\lambda_{B_{i}}$) of
$e_A$ ($e_B$) in the state 
$p_A$ ($p_B$), $i=1,2$, such that $p(\lambda_{A_{1}})+p(\lambda_{A_{2}})=p(\lambda_{B_{1}})+p(\lambda_{B_{2}})=1$, and $p(\lambda_{A_{i}B_{j}})=p(\lambda_{A_i})p(\lambda_{B_{j}})$, where $\lambda_{A_{i}B_{j}}$, $i,j=1,2$ are the outcomes of $e$ in the state $p$. 
\end{theorem}
From Th. \ref{th2} follows that, if the probabilities $p(\lambda_{A_{i}B_{j}})$ do not factorize, then only three possibilities exist: (i) the state $p$ is not a product state; (ii) the measurement $e$ is not a product measurement; (iii) both $p$ is not a product state, and $e$ is not a product measurement.

Let us consider the coincidence measurements $e_{AB}$, $e_{AB'}$, $e_{A'B}$ and $e_{A'B'}$ introduced above, together with their outcomes and probabilities in the state $p$. 
\begin{definition} \label{marginallaw}
We say that a set of experimental data ${\mathcal S}_{p}(AB)$ collected on the measurement $e_{AB}$ satisfies the `marginal distribution law' if, for every $i=1,2$,
\begin{eqnarray}
\sum_{j=1,2} p(\lambda_{A_{i}B_{j}})=\sum_{j=1,2} p(\lambda_{A_{i}B'_{j}}) \\
\sum_{j=1,2} p(\lambda_{A_{j}B_{i}})=\sum_{j=1,2} p(\lambda_{A'_{j}B_{1}}) 
\end{eqnarray}
We say that the marginal distribution law is satisfied in a Bell test if it is satisfied by all measurements $e_{AB}$, $e_{AB'}$, $e_{A'B}$ and $e_{A'B'}$.
\end{definition}
\begin{theorem} \label{productmarginal}
Let $e$ be a product measurement. Then, the marginal distribution law is satisfied by $e$.
\end{theorem}
\begin{theorem} \label{measuremententanglement}
If no measurement among $e_{AB}$, $e_{AB'}$, $e_{A'B}$ and $e_{A'B'}$ satisfy the marginal distribution law, then at least two measurements are entangled.
\end{theorem}
The latter provide a sharp and complete description of the structural situation: `entanglement is a relational property of states and measurements'. If Th. \ref{th2} is not satisfied by a set of data collected in a single measurement, then one can transfer all the entanglement in the state, or in the measurement, or in both. Theorem \ref{measuremententanglement} then shows that, if the marginal distribution law is violated, no more than two measurements can be products. The main consequence is that, if a set of experimental data violate both Bell's inequalities and the marginal distribution law, then a quantum-mechanical representation in the Hilbert space ${\mathbb C}^2\otimes{\mathbb C}^2$ cannot be worked out which satisfies the data and where only the initial state is entangled while all measurements are products. 

Finally, we remind a technical result on the violation of the CHSH inequalitiy with product measurements. The standard quantum inequality $\Delta \le 2\sqrt{2}$ holds for product measurements and is called `Tsirelson's bound' \cite{tsirelson80}. This is typically considered the maximal violation of Bell's inequalities that is allowed by quantum theory, and will be discussed in the following sections.

\subsection{Entanglement and mixtures\label{mixtures}}
We have proved in Sec. \ref{qmath}, Th. \ref{th2}, that entanglement in a state-measurement situation can be traced by investigating whether the probabilities factorize, in the case of pure states. We show in this section that this does not hold any longer in the case of mixtures and provide a criterion for the identification of entanglement in the latter case. To this end, let us consider an entity $S$ prepared in the mixture $m$ of the pure product states $r_1$, $r_2$, \ldots, represented by the unit vectors 
$|r_{A_1}\rangle \otimes |r_{B_1}\rangle$,  
$|r_{A_2}\rangle \otimes |r_{B_2}\rangle$, \ldots, with weights $w_{1}$, $w_{2}$, \ldots, respectively ($w_i \ge 0$, $\sum_{i} w_i=1$). The mixture $m$ is thus represented by the density operator 
\begin{equation}
\rho=\sum_{i} w_{i} |r_{A_i}\rangle\langle r_{A_i}|\otimes |r_{B_i}\rangle\langle r_{B_i}|
\end{equation}
Suppose that the measurement 
$e_{AB}$ 
is a product measurement, represented by the self-adjoint operator 
${\mathcal E}_{AB}={\mathcal E}_{A} \otimes {\mathcal E}_{B}$, 
with spectral family on the ON basis $\{ |p_{A_{1}B_{1}}\rangle,  |p_{A_{1}B_{2}}\rangle,|p_{A_{2}B_{1}}\rangle, |p_{A_{2}B_{2}}\rangle \}$, where $|p_{A_{i}B_{j}}\rangle=|p_{A_{i}}\rangle \otimes |p_{B_{j}}\rangle$, $i,j=1,2$, while the spectral families of ${\mathcal E}_{A}$ and ${\mathcal E}_{B}$ are constructued on the ON bases $\{ |p_{A_{i}}\rangle\}_{i=1,2}$ and $\{ |p_{B_{j}}\rangle\}_{j=1,2}$, respectively. The probability $p_{m}(\lambda_{A_{1}B_{1}})$ that the outcome $\lambda_{A_{1}B_{1}}$ is obtained when $e_{AB}$ is performed on $S$ in the mixture $m$ is
\begin{equation}
p_{m}(\lambda_{A_{1}B_{1}})=Tr [\rho |p_{A_{1}B_{1}}\rangle\langle p_{A_{1}B_{1}}|]=\sum_{i} w_i |\langle p_{A_{1}}|r_{A_i}\rangle|^{2}|\langle p_{B_{1}}|r_{B_i}\rangle|^{2}=\sum_{i} w_i p_{i}(\lambda_{A_{1}}) p_{i}(\lambda_{B_{1}})
\end{equation}
where $p_{i}(\lambda_{A_1})$ and $p_{i}(\lambda_{B_{1}})$ are the probabilities for the sub-measurements $e_{A}$ and $e_{B}$, respectively. Analogous formulas hold for the other outcomes, as follows
\begin{eqnarray}
p_{m}(\lambda_{A_{1}B_{2}})=\sum_{i} w_i p_{i}(\lambda_{A_{1}}) p_{i}(\lambda_{B_{2}}) \\
p_{m}(\lambda_{A_{2}B_{1}})=\sum_{i} w_i p_{i}(\lambda_{A_{2}}) p_{i}(\lambda_{B_{1}}) \\
p_{m}(\lambda_{A_{2}B_{2}})=\sum_{i} w_i p_{i}(\lambda_{A_{2}}) p_{i}(\lambda_{B_{2}})
\end{eqnarray}
Hence, if we start with the numbers $p_{m}(\lambda_{A_{i}B_{j}})$, $i,j=1,2$, then it is not possible in general to prove that no mixture of product states that exist that gives rise to these numbers. 

The violation of the marginal distribution law remains however a criterion for the presence of genuine entanglement. Indeed, let $m$ be a mixture represented by the density operator $\rho$, and let us suppose that, e.g., the measurement $e_{AB}$, is a product measurement, represented by the self-adjoint operator 
${\mathcal E}_{A} \otimes {\mathcal E}_{B}$, with spectral family on the ON basis $\{ |p_{A_{1}B_{1}}\rangle,  |p_{A_{1}B_{2}}\rangle,|p_{A_{2}B_{1}}\rangle, |p_{A_{2}B_{2}}\rangle \}$, where $|p_{A_{i}B_{j}}\rangle=|p_{A_{i}}\rangle \otimes |p_{B_{j}}\rangle$, $i,j=1,2$. We have
\begin{eqnarray}
p_{m}(\lambda_{A_{1}B_{1}})+p_{m}(\lambda_{A_{1}B_{2}})
&=&Tr[\rho |p_{A_{1}B_{1}}\rangle\langle p_{A_{1}B_{1}}|]
+Tr[\rho |p_{A_{1}B_{2}}\rangle\langle p_{A_{1}B_{2}}|] \nonumber \\
&=&p_{m}(\lambda_{A_{1}})=p_{m}(\lambda_{A_{1}B'_{1}})+p_{m}(\lambda_{A_{1}B'_{2}})
\end{eqnarray}
This means that the identification of a violation of the marginal law remains an indication of the presence of genuine entanglement.

Let us now consider Bell's inequalities in case of a mixture of product states and product measurements $e_{AB}$, $e_{AB'}$, $e_{A'B}$ and $e_{A'B'}$. They are respectively represented by the `expectation value operators' ${\mathcal E}_{AB}={\mathcal E}_{A} \otimes {\mathcal E}_{B}$, ${\mathcal E}_{AB'}={\mathcal E}_{A} \otimes {\mathcal E}_{B'}$, ${\mathcal E}_{A'B}={\mathcal E}_{A'} \otimes {\mathcal E}_{B}$ and ${\mathcal E}_{A'B'}={\mathcal E}_{A'} \otimes {\mathcal E}_{B'}$, where ${\mathcal E}_{A}=|p_{A_1}\rangle\langle p_{A_1}|-|p_{A_2}\rangle\langle p_{A_2}|$, \ldots, 
${\mathcal E}_{B'}=|p_{B'_1}\rangle\langle p_{B'_1}|-|p_{B'_2}\rangle\langle p_{B'_2}|$. We have
\begin{eqnarray}
{\mathcal E}_{AB'}+{\mathcal E}_{A'B'}=({\mathcal E}_{A}+{\mathcal E}_{A'}) \otimes {\mathcal E}_{B'} \\
{\mathcal E}_{A'B}-{\mathcal E}_{AB}=({\mathcal E}_{A'}-{\mathcal E}_{A}) \otimes {\mathcal E}_{B}
\end{eqnarray}
Hence, 
one of the Bell operators is given by
\begin{eqnarray}
B&=& {\mathcal E}_{AB'}+{\mathcal E}_{A'B'}+{\mathcal E}_{A'B}-{\mathcal E}_{AB}\nonumber \\
&=& ({\mathcal E}_{A}+{\mathcal E}_{A'}) \otimes {\mathcal E}_{B'}+({\mathcal E}_{A'}-{\mathcal E}_{A}) \otimes {\mathcal E}_{B} 
\end{eqnarray}
Suppose we consider a product state $p$ represented by the unit vector $|p\rangle = |r_A\rangle \otimes |r_B\rangle$. The factor $\Delta$ in the CHSH inequality is, in this case,
\begin{eqnarray}
\Delta&=&\langle p|B|p\rangle=\langle r_A| {\mathcal E}_{A} | r_A \rangle \langle r_B |{\mathcal E}_{B'}| r_B \rangle \nonumber  + 
\langle r_A | {\mathcal E}_{A'} | r_A \rangle \langle r_B |{\mathcal E}_{B'} | r_B \rangle  \nonumber \nonumber \\
&&+\langle r_A | {\mathcal E}_{A'}| r_A \rangle \langle r_B | {\mathcal E}_{B} | r_B \rangle- \langle r_A | {\mathcal E}_{A}| r_A \rangle \langle r_B | {\mathcal E}_{B} | r_B \rangle \label{lemmaprod}
\end{eqnarray}
Let us consider the following mathematical result.
\newtheorem{lemma}{Lemma}
\begin{lemma} \label{l1}
If $x$, $x'$, $y$ and $y'$ are real numbers such that $-1\le x, x', y , y\le +1$ and $\Delta=x'y' + x'y + xy' - xy$, then $-2\le \Delta \le +2$.
\begin{proof}
Since $\Delta$ is linear in all the variables $x$, $x'$, $y$, $y'$, it must take on its maximum and minimum values at the corners of the domain of this quadruple of variables, that is, where each of $x$, $x'$, $y$, $y'$ is +1 or -1. Hence at these corners $\Delta$ can only be an integer between -4 and +4. But $\Delta$ can be rewritten as $(x + x')(y + y') - 2xy$, and the two quantities in parentheses can only be 0, 2, or -2, while the last term can only be -2 or +2, so that $\Delta$ cannot equal -3, +3, -4, or +4 at the corners.
\end{proof}
\end{lemma}
One can verify at once that Eq. (\ref{lemmaprod}) satisfies Lemma \ref{l1}, i.e. $-2 \le \langle p|B|p\rangle \le +2$. Hence, the CHSH inequality holds whenever $p$ is a product state.

Let us investigate whether we can prove that Bell's inequalities are satisfied also when $p$ is a mixture of product states. Hence, let $m$ be a mixture of the pure product states $r_1$, $r_2$, \ldots, represented by the unit vectors $|r_{A_1}\rangle \otimes |r_{B_1}\rangle$,  $|r_{A_2}\rangle \otimes |r_{B_2}\rangle$, \ldots, with weights $w_{1}$, $w_{2}$, \ldots, respectively, so that $m$ is represented by the density operator $\rho=\sum_{i} w_{i} |r_{A_i}\rangle\langle r_{A_i}|\otimes |r_{B_i}\rangle\langle r_{B_i}|$. We have, by using Eq. (\ref{lemmaprod}),
\begin{eqnarray}
\Delta&=&Tr [\rho B]=Tr [\sum_{i}w_i |r_{A_i}\rangle\langle r_{A_i}|\otimes |r_{B_i}\rangle\langle r_{B_i}|B] \nonumber \\
&=&\sum_{i} w_{i} \Big ( \langle r_{A_i} | {\mathcal E}_{A} | r_{A_i} \rangle \langle r_{B_i} |{\mathcal E}_{B'}| r_{B_i} \rangle+ 
\langle r_{A_i} | {\mathcal E}_{A'} | r_{A_i} \rangle \langle r_{B_i} |{\mathcal E}_{B'} | r_{B_i} \rangle  \nonumber \nonumber \\
&&+\langle r_{A_i} | {\mathcal E}_{A'}| r_{A_i} \rangle \langle r_{B_i} | {\mathcal E}_{B} | r_{B_i} \rangle - \langle r_{A_i} | {\mathcal E}_{A}| r_{A_i} \rangle \langle r_{B_i} | {\mathcal E}_{B} | r_{B_i} \rangle \Big )
\end{eqnarray}
If we now put, for every $i$, 
\begin{eqnarray}
\delta_i&=&\langle r_{A_i} | {\mathcal E}_{A} | r_{A_i} \rangle \langle r_{B_i} |{\mathcal E}_{B'}| r_{B_i} \rangle + 
\langle r_{A_i} | {\mathcal E}_{A'} | r_{A_i} \rangle \langle r_{B_i} |{\mathcal E}_{B'} | r_{B_i} \rangle  \nonumber \\
&&+\langle r_{A_i} | {\mathcal E}_{A'}| r_{A_i} \rangle \langle r_{B_i} | {\mathcal E}_{B} | r_{B_i} \rangle \nonumber - \langle r_{A_i} | {\mathcal E}_{A}| r_{A_i} \rangle \langle r_{B_i} | {\mathcal E}_{B} | r_{B_i} \rangle
\end{eqnarray}
we get from Lemma \ref{l1} that, for every $i$, $-2 \le \delta_i \le +2$. Then, we can prove that $-2 \le \Delta=\sum_{i} w_i \delta_i \le +2$. Indeed, if, for every $i$, $\delta_i=+2$, we have $\sum_{i}w_i\delta_i=2\sum_{i}w_i=+2$. Analogously, if, for every $i$, $\delta_i=-2$, we have $\sum_{i}w_i\delta_i=-2\sum_{i}w_i=-2$. Since now $\sum_{i} w_i \delta_i$ is a convex combination of $\delta_i$, with weights $w_i$, its value lies in the convex set of numbers with extremal points -2 and +2, hence in the interval $[-2, +2]$, as maintained above. This proves that the CHSH inequality is satisfied when the situation is such that we have product measurements and a mixture of product states.

Summing up the results obtained in this section, we can say that the structural situation is the following.

(i) If product measurements are performed, the marginal distribution law holds whenever the state is a pure product state or a mixture of product states. This entails that there is genuine entanglement when the marginal distribution law is violated, independent of the state of the entity.

(ii) When the marginal distribution law is satisfied and Bell's inequalities are violated, we have genuine entanglement. Indeed, Bell's inequalities hold also for a mixture of product states. In this case, the validiy of the marginal distribution law entails that all the entanglement can be pushed into the state.

(iii) When the marginal distribution law is satisfied, and the state is pure, we have genuine entanglement when the probabilities do not factorize. Indeed, for a pure product state and the marginal law satisfied, hence product measurements, the probabilities factorize.

It remains to investigate whether we can find more direct criteria for genuine entanglement allowing states to be mixtures, without the need to recur to the violation of Bell's inequalities. Indeed, entanglement exists which does not violate Bell's inequalities, hence the latter violation is only a sufficient condition.

\section{Quantum realization of a nonlocal non-marginal box\label{nonlocalnonmarginalbox}}
In this section, we elaborate a quantum model in Hilbert space for an entity violating both Tsirelson's bound and the marginal distribution law. This case study manifestly reveals that the costruction of a quantum model is allowed by entangled measurements. Experimental realizations of this situation in physical and cognitive situations can be found in \cite{1asQI2013,2asQI2013}.  

Let $S$ be an entity prepared in the pure entangled state $p$. Bell-type measurements are defined as usual. The measurement $e_{AB}$ has the outcomes $\lambda_{A_{1}B_{1}}$, $\lambda_{A_{1}B_{2}}$, $\lambda_{A_{2}B_{1}}$ and $\lambda_{A_{2}B_{2}}$, and the final states, $p_{A_{1}B_{1}}$, $p_{A_{1}B_{2}}$, $p_{A_{2}B_{1}}$ and $p_{A_{2}B_{2}}$. The measurement $e_{AB'}$ has the outcomes $\lambda_{A_{1}B'_{1}}$, $\lambda_{A_{1}B'_{2}}$, $\lambda_{A_{2}B'_{1}}$ and $\lambda_{A_{2}B'_{2}}$, and the final states, $p_{A_{1}B'_{1}}$, $p_{A_{1}B'_{2}}$, $p_{A_{2}B'_{1}}$ and $p_{A_{2}B'_{2}}$. The measurement $e_{A'B}$ has four outcomes $\lambda_{A'_{1}B_{1}}$, $\lambda_{A'_{1}B_{2}}$, $\lambda_{A'_{2}B_{1}}$ and $\lambda_{A'_{2}B_{2}}$, and the final states $p_{A'_{1}B_{1}}$, $p_{A'_{1}B_{2}}$, $p_{A'_{2}B_{1}}$ and $p_{A'_{2}B_{2}}$. The measurement $e_{A'B'}$ has the outcomes $\lambda_{A'_{1}B'_{1}}$, $\lambda_{A'_{1}B'_{2}}$, $\lambda_{A'_{2}B'_{1}}$ and $\lambda_{A'_{2}B'_{2}}$, and the final states $p_{A'_{1}B'_{1}}$, $p_{A'_{1}B'_{2}}$, $p_{A'_{2}B'_{1}}$ and $p_{A'_{2}B'_{2}}$.

To work out a quantum-mechanical model for the latter situation in the Hilbert space ${\mathbb C}^2\otimes{\mathbb C}^2$, considering it canonical isomorphic with ${\mathbb C}^4$, we represent the entangled state $p$ by the unit vector $|p\rangle=|0, \sqrt{0.5}e^{i\alpha}, \sqrt{0.5}e^{i\beta}, 0\rangle$. The measurement $e_{AB}$ is represented by the ON (canonical) basis $|p_{A_{1}B_{1}}\rangle=|1, 0, 0, 0\rangle$, $|p_{A_{1}B_{2}}\rangle=|0, 1, 0, 0\rangle$, $|p_{A_{2}B_{1}}\rangle=|0, 0, 1, 0\rangle$, $|p_{A_{2}B_{2}}\rangle=|0, 0, 0, 1\rangle$, and hence the probabilities of the outcomes $\lambda_{A_{i}B_{j}}$ of $e_{AB}$ in the state $p$ are given by $p(\lambda_{A_{1}B_{1}})=|\langle p_{A_{1}B_{1}}|p\rangle|^2=0$, $p(\lambda_{A_{1}B_{2}})=|\langle p_{A_{1}B_{2}}|p\rangle|^2=0.5$, $p(\lambda_{A_{2}B_{1}})=|\langle p_{A_{2}B_{1}}|p\rangle|^2=0.5$, $p(\lambda_{A_{2}B_{2}})=|\langle p_{A_{2}B_{2}}|p\rangle|^2=0$.

The measurement $e_{AB'}$ is represented by the ON basis
$|p_{A_{1}B'_{1}}\rangle=|0, \sqrt{0.5}e^{i\alpha}, \sqrt{0.5}e^{i\beta}, 0\rangle$, $|p_{A_{1}B'_{2}}\rangle=|0, \sqrt{0.5}e^{i\alpha}, -\sqrt{0.5}e^{i\beta}, 0\rangle$,
$|p_{A_{2}B'_{1}}\rangle=| 1, 0, 0, 0\rangle$, $|p_{A_{2}B'_{2}}\rangle=|0, 0, 0, 1\rangle$, and the probabilities of the outcomes $\lambda_{A_{i}B'_{j}}$ of $e_{AB'}$ in the state $p$ are given by
 $p(\lambda_{A_{1}B'_{1}})=|\langle p_{A_{1}B'_{1}}|p\rangle|^2=1$, 
 $p(\lambda_{A_{1}B'_{2}})=|\langle p_{A_{1}B'_{2}}|p\rangle|^2=0$, 
 $p(\lambda_{A_{2}B'_{1}})=|\langle p_{A_{2}B'_{1}}|p\rangle|^2=0$,  
 $p(\lambda_{A_{2}B'_{2}})=|\langle p_{A_{2}B'_{2}}|p\rangle|^2=0$. 

The measurement $e_{A'B}$ is represented by the ON basis  $|p_{A'_{1}B_{1}}\rangle=|0, \sqrt{0.5}e^{i\alpha}, \sqrt{0.5}e^{i\beta}, 0\rangle$,
$|p_{A'_{1}B_{2}}\rangle= | 1, 0, 0, 0\rangle$, $|p_{A'_{2}B_{1}}\rangle= |0, \sqrt{0.5}e^{i\alpha}, -\sqrt{0.5}e^{i\beta}, 0\rangle$, $|p_{A'_{2}B_{2}}\rangle= |0, 0, 0, 1\rangle$, which entails probability 1 for the outcome $\lambda_{A'_{1}B_{1}}$ in the state $p$.

Finally, the measurement $e_{A'B'}$ is represented by the ON basis
$|p_{A'_{1}B'_{1}}\rangle=|0, \sqrt{0.5}e^{i\alpha}, \sqrt{0.5}e^{i\beta}, 0\rangle$, $|p_{A'_{1}B'_{2}}\rangle= | 1, 0, 0, 0\rangle$,  
$|p_{A'_{2}B'_{1}}\rangle= |0, 0, 0, 1\rangle$,  $|p_{A'_{2}B'_{2}}\rangle= |0, \sqrt{0.5}e^{i\alpha}, -\sqrt{0.5}e^{i\beta}, 0\rangle$, which entails probability 1 for the outcome $\lambda_{A'_{1}B'_{1}}$ in the state $p$.

Let us now explicitly construct the self-adjoint operators representing the measurements $e_{AB}$, $e_{AB'}$, $e_{A'B}$ and $e_{A'B'}$. They are respectively given by  
\begin{eqnarray}
&{\mathcal E}_{AB}=\sum_{i,j=1}^{2}\lambda_{A_{i}B_{j}}|p_{A_{i}B_{j}}\rangle \langle p_{A_{i}B_{j}}|
\label{1}
\\
&{\mathcal E}_{AB'}=\sum_{i,j=1}^{2}\lambda_{A_{i}B'_{j}}|p_{A_{i}B'_{j}}\rangle \langle p_{A_{i}B'_{j}}| 
\label{2}
\\
&{\mathcal E}_{A'B}=\sum_{i,j=1}^{2}\lambda_{A'_{i}B_{j}}|p_{A'_{i}B_{j}}\rangle \langle p_{A'_{i}B_{j}}|
\label{3}
\\
&{\mathcal E}_{A'B'}=\sum_{i,j=1}^{2}\lambda_{A'_{i}B'_{j}}|p_{A'_{i}B'_{j}}\rangle \langle p_{A'_{i}B'_{j}}|
\label{4}
\end{eqnarray}
The self-adjoint operators corresponding to measuring the expectation values are obtained by putting 
$\lambda_{A_{i}B_{i}}=\lambda_{A_{i}B'_{i}}=\lambda_{A'_{i}B_{i}}=\lambda_{A'_{i}B'_{i}}=+1$, $i=1,2$ and $\lambda_{A_{i}B_{j}}=\lambda_{A_{i}B'_{j}}=\lambda_{A'_{i}B_{j}}=\lambda_{A'_{i}B'_{j}}=-1$, $i,j=1,2;i \ne j$. If we now insert these values into Eqs. (\ref{1})--(\ref{4}) and define 
one of the `Bell operators' as
\begin{eqnarray}
B={\mathcal E}_{AB'}+{\mathcal E}_{A'B}+{\mathcal E}_{AB'}-{\mathcal E}_{AB} \nonumber\\
=\left( \begin{array}{cccc}
0 & 0 & 0 & 0 \\
0 & 2 & 2e^{i(\alpha-\beta)} & 0 \\
0 & 2e^{-i(\alpha-\beta)} & 2 & 0 \\
0 & 0 & 0 & 0
\end{array} \right)
\end{eqnarray}
and its expectation value in the entangled state $p$,  
we get $\Delta=\langle p |B|p\rangle=4$ in the CHSH inequality.

We add some conclusive remarks that are discussed in detail in \cite{1asQI2013,2asQI2013}. The measurement $e_{AB}$ is a product measurement, since it has the product states represented by the vectors in the canonical basis of ${\mathbb C}^{4}$ as final states. Hence, $e_{AB}$ `destroys' the initial entanglement to arrive at a situation of a product state. The measurements $e_{AB'}$, $e_{A'B}$ and $e_{A'B'}$ are instead entangled measurements, since they are represented by spectral families constructed on entangled states (Th. \ref{th1}).  We finally observe that the marginal ditribution law is violated. Indeed, we have, e.g., $0.5=p(\lambda_{A_{1}B_{1}})+p(\lambda_{A_{1}B_{2}}) \ne p(\lambda_{{A_1}B'_{1}})+p(\lambda_{{A_1}B'_{2}})=1$. Since then the situation above violates Bell's inequalities beyond Tsirelson's bound, we can say that we have an example of a `nonlocal non-marginal box situation 2', if we follow the classification in Sec. \ref{intro}. The locution `nonlocal non-marginal box situation 1' has instead been used to denote a situation violating the marginal distribution law, but not Tsirelson's bound \cite{asIQSA2012,1asQI2013,2asQI2013}. It seems that situations of this kind have been observed in Bell-type experiments on microscopic quantum particles, where they have been classified as `anomalies' \cite{aspect1982,ak}. We are elaborating an explanation of these anomalies in terms of entangled measurements within our quantum-theoretic framework. From a quantum foundational point of view, such an explanation would (i) constitute a breakthrough toward understanding the mechanism of entanglement, (ii) shed new light into the so-called `no-signaling problem'.

The mathematical description presented here, modeling physical and cognitive experimental examples and in other papers \cite{1asQI2013,2asQI2013} are relevant, in our opinion, because they explictly show that a quantum model in Hilbert space can be elaborated also for a situation going beyond Tsirelson's bound, if one introduces entangled measurements. This reveals that the violation of Bell's inequalities is `not limited by Tsirelson's bound' and can even be maximal, as we will see in the next section too.

\section{Quantum realization of a nonlocal box\label{nonlocalbox}}
In this section, we provide a quantum Hilbert space modeling for an entity which maximally violate Bell's inequalities, i.e. with value 4, but satisfies the marginal distribution law. In physics, a system that behaves in this way is called a `nonlocal box' \cite{rp94}. We will see that such a system exhibits the typical symmetry which gives rise to the marginal distribution law being valid in quantum theory. Concrete experimental realizations of this situation can be found in \cite{1asQI2013}.   

We consider four measurements $f_{AB}$, $f_{AB'}$, $f_{A'B}$ and $f_{A'B'}$, with outcomes $\mu_{A_{i}B_{j}}$, \ldots, and $\mu_{A'_{i}B'_{j}}$, $i,j=1,2$, respectively, and a composite entity $S$ in the mixture $m$ of the pure entangled states $p$ and $q$, represented by the unit vectors $|p\rangle=|0,\sqrt{0.5}e^{i\alpha},0.5e^{i\beta},0 \rangle$ and $|q\rangle=|0,\sqrt{0.5}e^{i\alpha},-0.5e^{i\beta},0 \rangle$, respectively, with equal weights. Thus, $m$ is represented by the density operator $\rho=0.5|p\rangle\langle p|+0.5|q\rangle\langle q|$. 

The first measurement $f_{AB}$ is represented by the ON basis 
$|r_{A_{1}B_{1}}\rangle=|1, 0, 0, 0\rangle$,
$|r_{A_{1}B_{2}}\rangle=|0, 1, 0, 0\rangle$, 
$|r_{A_{2}B_{1}}\rangle=|0, 0, 1, 0\rangle$, 
$|r_{A_{2}B_{2}}\rangle=|0, 0, 0, 1\rangle$, which gives rise to the self-adjoint operator
\begin{equation}
{\mathcal F}_{AB}=\sum_{i,j}\mu_{A_{i}B_{j}} |r_{A_{i}B_{j}}\rangle\langle r_{A_{i}B_{j}}|
\end{equation}

By applying L\"{u}ders' rule, we can now calculate the density operator representing the final state of the entity $S$ after the measurement $f_{AB}$. This gives
\begin{equation}
\rho_{AB}=\sum_{i,j=1}^{2} |r_{A_{i}B_{j}}\rangle\langle r_{A_{i}B_{j}}|\rho|r_{A_{i}B_{j}}\rangle\langle r_{A_{i}B_{j}}|=\rho
\end{equation}
as one can easily verify. This means that the nonselective measurement $f_{AB}$ leaves the state $m$ unchanged or, equivalently, the marginal distribution law holds, in this case.

The second measurement $f_{AB'}$ is represented by the ON basis $|r_{A_{1}B'_{1}}\rangle=|0, \sqrt{0.5}e^{i\alpha}, \sqrt{0.5}e^{i\beta}, 0\rangle$, 
$|r_{A_{1}B'_{2}}\rangle= | 1, 0, 0, 0\rangle$, 
$|r_{A_{2}B'_{1}}\rangle= |0, 0, 0, 1\rangle$, 
$|r_{A_{2}B'_{2}}\rangle= |0, \sqrt{0.5}e^{i\alpha}, -\sqrt{0.5}e^{i\beta}, 0\rangle$,
which gives rise to a self-adjoint operator
\begin{equation}
{\mathcal F}_{AB'}=\sum_{i,j}\mu_{A_{i}B'_{j}} |r_{A_{i}B'_{j}}\rangle\langle r_{A_{i}B'_{j}}|
\end{equation}

By applying L\"{u}ders' rule, we can again calculate the density operator representing the final state of the vessels of water after $f_{AB'}$. This gives
\begin{equation}
\rho_{AB'}=\sum_{i,j=1}^{2} |r_{A_{i}B'_{j}}\rangle\langle r_{A_{i}B'_{j}}|\rho|r_{A_{i}B'_{j}}\rangle\langle r_{A_{i}B'_{j}}|=\rho
\end{equation}
Also in this case, the nonselective measurement $f_{AB'}$ leaves the state $m$ unchanged. If we consider the experimental realization in \cite{1asQI2013} of the nonlocal box situation, we can see that we can represent the measurements $f_{A'B}$ and $f_{A'B'}$ by the same self--adjoint operators as the one representing $f_{AB'}$.
Also in these cases we obviously get that the density operators after applying L\"{u}ders' rule remain the same. This implies that the marginal distribution law is always satisfied. 

Let us now evaluate the expectation values corresponding to the four measurements above in the mixed state $m$ and insert them into the CHSH inequality. The expectation value operators for this version are given by
\begin{eqnarray}
{\mathcal F}_{AB}=
\left( \begin{array}{cccc}
1 & 0 & 0 & 0 \\
0 & -1 & 0 & 0 \\
0 & 0 & -1 & 0 \\
0 & 0 & 0 & 1
\end{array} \right)
\\
{\mathcal F}_{AB'}={\mathcal F}_{A'B}={\mathcal F}_{A'B'}=
\left( \begin{array}{cccc}
-1 & 0 & 0 & 0 \\
0 & 1 & 0 & 0 \\
0 & 0 & 1 & 0 \\
0 & 0 & 0 & -1
\end{array} \right)
\end{eqnarray}
Hence,  
our Bell operator is given by
\begin{equation}
B={\mathcal F}_{AB'}+{\mathcal F}_{A'B}+{\mathcal F}_{A'B'}-{\mathcal F}_{AB}=
\left( \begin{array}{cccc}
-4 & 0 & 0 & 0 \\
0 & 4 & 0 & 0 \\
0 & 0 & 4 & 0 \\
0 & 0 & 0 & -4
\end{array} \right)
\end{equation}
This gives $\Delta=Tr \rho B = 4$ in the CHSH inequality, which shows that Bell's inequalities are maximally violated in the mixture $m$. Following our classification scheme in Sec. \ref{intro}, we can can regard our quantum model for the nonlocal box above as an example of a `nonlocal box situation'. 

In the foundations of quantum theory, the possibility of constructing a quantum representation for a nonlocal box is usually maintained to be forbidden by quantum laws, i.e. Tsirelson's bound. We have shown here that
such a quantum representation
can indeed be elaborated, once entangled measurements are taken into account.

\section{Conclusions\label{conclusions}}
We have presented a quantum-theoretic modeling in Hilbert space for the description of the entanglement that characterizes situations experimentally violating Bell's inequalities. We have shown that different types of quantum models can be constructed, in addition to the
`customary quantum situation', depending on the behavior with respect to (i) the marginal distribution law, (ii) Tsirelson's bound. Moreover, entangled measurements provide an operational and technical resource for dealing with situations that are typically considered `beyond the customary quantum situation'. This scheme has been also extended to quantum mixtures, attaining some nontrivial conclusions. 

The perspective above is completely general, for it enables detection and structural description of the entanglement that is present in any kind of composite entity (microscopic particles, combinations of concepts, financial assets, biological aggregates, etc.), once experimental tests are defined giving rise to the scheme necessary to formulate Bell's inequalities. We also believe our scheme to be valuable for the study of quantum foundational problems in a more general way. Indeed, the introduction of entangled measurements introduces a new understanding of the entanglement dynamics in Bell-type experiments, also on microscopic quantum particles and nonlocal boxes. The realization of an experimental nonlocal box may have a deep impact on the technologies employed in quantum information to detect, measure and preserve entanglement.

\end{document}